\documentclass[11pt,a4paper,eps]{article}
\usepackage{graphicx,amssymb}
\usepackage{amsmath}
\usepackage{graphicx}
\usepackage{amsfonts}
\usepackage{amssymb}
\begin{document}
\pagestyle{empty}

\begin{flushright}
\begin{tabular}{c}
hep-th/0304090
\end{tabular}
\end{flushright}

\vspace{1.5cm}

\begin{center}
{\LARGE Cornwall-Jackiw-Tomboulis effective potential for
canonical noncommutative field theories}
\end{center}

\vskip1.5 cm

\begin{center}
\textbf{Gianluca MANDANICI} \vskip0.3 cm

\textit{Dipartimento di Fisica, Universit\`{a} di Roma ``La
Sapienza'', }

\textit{P.le A. Moro 2, 00185 Roma, Italy}
\end{center}

\vspace{1.5cm}

\begin{center}
\textbf{ABSTRACT}
\end{center}

{\leftskip=0.6in \rightskip=0.6in\noindent}

We apply the Cornwall-Jackiw-Tomboulis (CJT) formalism to the
scalar $\lambda \varphi^{4}$ theory in canonical-noncommutative
spacetime. We construct the CJT effective potential and the gap
equation for general values of the noncommutative parameter
$\theta_{\mu\nu}$. We observe that under the hypothesis of
translational invariance, which is assumed in the effective
potential construction, differently from the commutative case
($\theta_{\mu\nu}= 0$), the renormalizability of the gap equation
is incompatible with the renormalizability of the effective
potential. We argue that our result, is consistent with previous
studies suggesting that a uniform ordered phase would be
inconsistent with the infrared structure of canonical
noncommutative theories.

\newpage\baselineskip16pt plus .5pt minus .5pt \pagenumbering{arabic}
\pagestyle{plain}

\section{Introduction}

Quantum field theories on canonical-noncommutative spacetime have
been introduced in \cite{DFR,filk}. Since then, noncommutative
field theories have been extensively studied in literature (for a
review see \cite{ND,RSz}) especially for their connection with
string theory \cite{cds,SW,A&A}. A key feature of canonical
noncommutative theories is the so-called IR/UV mixing which
connects the high-energy degrees of freedom of the theory with the
low-energy degrees of freedom \cite{mrs,MST,RS}. IR/UV mixing is a
direct consequence of the nonvanishing commutations relation.
Besides having strong implication for the phenomenological
predictions (see e.g. \cite{adns,aad,amk}), IR/UV mixing renders
the renormalizability and the infrared structure of noncommutative
field theories highly non-trivial (and interconnected) issues,
even in the case of massive theories. Despite these mentioned
infrared problems, in the $\lambda\varphi^{4}$-theory case,
one-loop renormalization was explicitly carried out in \cite{mrs}.
In \cite{ABK,MJ}\ two-loop renormalization was obtained, and in
\cite{gripi}, using the Polchinski method \cite{polchi},
renormalization was claimed to all orders of the perturbative
expansion. The Polchinski method was also adopted in the
discussion of the renormalizability of the $O(N)$-symmetric scalar
theory reported in \cite{sarkar}, where it was also argued that in
order to achieve renormalizability of the ordered phase it would
be necessary to relax the hypothesis of translational invariance
of the vacuum. The idea that noncommutative scalar theory exhibits
a transition to a non-uniform phase was first considered in some
detail in \cite{guso} where, for the ordered phase, a ``stripe
phase'' was proposed. Recently numerical studies provided support
for this hypothesis in lower dimensional cases \cite{BHN,AC,BHN2}.
Problems with the renormalization in the translationally invariant
ordered phase where also encountered in \cite{CK,ABS}\ and
\cite{RR} (however also see \cite{SSBS,GGPR}).

In this paper we want to investigate these issues pertaining to
the translationally invariant vacuum by means of the
Cornwall-Jackiw-Tomboulis (CJT) formalism \cite{CJT}. This
formalism has proven to be a powerful non-perturbative approach
for the study of phase transitions in QFT in commutative spacetime
\cite{CJT,jacgac}, especially in those theories exhibiting severe
infrared problems such as thermal field theories (see e.g.
\cite{jacgac,gacpi,gac1}). It is therefore natural to consider the
application of the CJT formalism to canonical noncommutative field
theories, where severe infrared problems, as mentioned, are
present. We consider the scalar-$\lambda \varphi^{4}$ theory case.
We construct the CJT effective potential and the gap equation for
general values of the noncommutative parameter
$\theta_{\mu\nu}$. We observe that under the hypothesis
of translational invariance, which is assumed in the effective
potential construction, differently from the commutative case
($\theta_{\mu\nu}= 0$), the renormalizability of the gap equation
is incompatible with the renormalizability of the effective
potential. We argue that our result, is consistent with previous
studies suggesting that a uniform ordered phase would be
inconsistent with the infrared structure of canonical
noncommutative theories.

The paper is organized as follows. In Section-2 we review the CJT
formalism in the commutative scalar $\lambda \varphi^{4}$ theory
also discussing the bubble approximation. In Section-3 we apply
CJT formalism to the noncommutative scalar $\lambda \varphi^{4}$
theory. In Section-4 we calculate the gap equation and the
effective potential and discuss their renormalizability. Then, in
Section-5, we report our conclusions.

\section{CJT formalism in commutative spacetime}

In this section we briefly review the CJT formalism for the
scalar-$\lambda\varphi^{4}$ theory\footnote{In this section, as
well as in the rest of the paper, we will work in euclidean
spacetime.} the commutative case \cite{CJT, jacgac}. The starting
point is the definition of the partition function
\[
Z[J,K]=\exp W[J,K]=\int D\phi\exp\left\{  S\left(  \phi\right)
+\int dx^{4}J(x)\phi(x)+\frac{1}{2}\int
dy^{4}dx^{4}\phi(x)K(x,y)\phi(y)\right\},
\]
in which two sources $\ J(x)$ and $K(x,y)$ are introduced.

One defines also $\varphi(x)$ and $G(x,y)$ by the relations
\begin{align}
 \frac{\delta
W[J,K]}{\delta J(x)}& =\varphi(x), \label{a1} \\
\frac{\delta W[J,K]} {\delta K(x,y)}&=\frac{1}{2}\left\{
\varphi(x)\varphi(y)+G(x,y)\right\} \label{a2}.
\end{align}
Then one performs a Legendre transform of $W[J,K]$:
\begin{align}
\Gamma\lbrack\varphi(x),G(x,y)]  &  =W[J(x),K(x,y)]-\int dx^{4}\varphi
(x)J(x)-\frac{1}{2}\int dx^{4}dy^{4}\varphi(x)K(x,y)\varphi(y)+\nonumber\\
&  -\frac{1}{2}\int dx^{4}dy^{4}G(x,y)K(x,y), \label{gamma}%
\end{align}
which satisfies the relations
\begin{align*}
\frac{\delta\Gamma\lbrack\varphi,G]}{\delta\varphi(x)}  &
=-J(x)-\int
dy^{4}K(x,y)\varphi(y),\\
\frac{\delta\Gamma\lbrack\varphi,G]}{\delta G(x,y)}  &  =-\frac{1}{2}K(x,y).
\end{align*}

The conditions for vanishing sources $K(x,y)=0$ and $J(x)=0$ (the
physical point), corresponds to choices of $\varphi(x)$ and
$G(x,y)$ such that they are solutions of the stationarity
equations:
\begin{align}
\frac{\delta\Gamma\lbrack\varphi,G]}{\delta\varphi(x)}  &  =0,\label{peq}\\
\frac{\delta\Gamma\lbrack\varphi,G]}{\delta G(x,y)}  &  =0. \label{gap}%
\end{align}
From (\ref{a1}) and (\ref{a2}) we see that, at the physical point,
$\varphi(x)$ is the vacuum-expectation-value of $\phi$, while
$G(x,y)$ is the expectation value of the connected part of the two
point function:
\begin{align}
\varphi(x)|_{J=K=0}&=\langle0|\phi(x)|0\rangle, \\
G(x,y)|_{J=K=0}&=\langle0|\phi(x)\phi(y)|0\rangle-\langle0|\phi(x)|0\rangle\langle0|\phi(y)|0\rangle.
\end{align}

It can be shown \cite{CJT}\ that $\Gamma\lbrack\varphi,G]$ so
defined is the generating functional for the two-particle
irreducible(2PI) Green's functions, with propagator given by
$G(x,y)$ and vertices given by $S_{int}(\varphi;\phi)$, where
$S_{int}(\varphi;\phi)$ is obtained from $S(\varphi)$ by retaining
only cubic and higher $\varphi$-terms in the expression of
$S(\varphi+\phi)$.

\begin{figure}[h]
\begin{center}
\includegraphics[width=5cm]{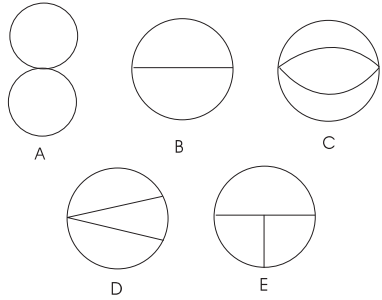}
\end{center}
\caption{2PI vacuum diagrams contributing to
$\Gamma^{2}(\varphi,G)$ up to the three loop level.}
\label{fig:vvdiagrams}%
\end{figure}

One can expand (\ref{gamma}) to obtain
\begin{equation}
\Gamma(\varphi,G)=S(\varphi)-\frac{1}{2}TrLnD_{0}^{-1}G+\frac{1}%
{2}Tr\left\{  D^{-1}G-1\right\}  +\Gamma^{2}(\varphi,G), \label{gammae}%
\end{equation}
where
\begin{align*}
D^{-1}(x,y)  &  =\frac{\delta^{2}S}{\delta\varphi(x)\delta\varphi(y)},\\
D_{0}^{-1}(x,y)  &  =D^{-1}|_{\lambda=0},
\end{align*}
and $\Gamma^{2}(\varphi,G)$ \ is the sum of the 2PI-vacuum
diagrams with vertices given by $S_{int}%
(\varphi;\phi)$ and propagators given by $G(x,y)$ (see
Fig.\ref{fig:vvdiagrams}).

\begin{figure}[h]
\begin{center}
\includegraphics[width=7cm]{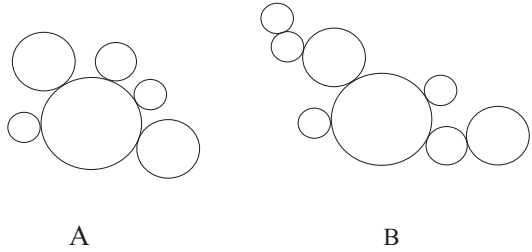}
\end{center}
\caption{Examples of bubble diagrams.}
\label{fig:bd}%
\end{figure}

Using (\ref{gammae}) the gap equation (\ref{gap}) may be rewritten
in the form
\begin{equation}
G^{-1}(x,y)=D^{-1}(x,y)+2\frac{\delta\Gamma^{2}(\varphi,G)}{\delta
G(x,y)}.
\label{gap2}%
\end{equation}
One can also recover the usual 1PI-effective action
$\Gamma(\varphi)$ simply evaluating $\Gamma\lbrack\varphi,G]$ at
nonvanishing $J(x)$ but at vanishing $K(x,y)$:
\begin{equation}
\Gamma_{1PI}(\varphi)=\Gamma[\varphi,G_{0}], \label{gamma1pi}
\end{equation}
where $G_{0}$ is solution of the gap equation (\ref{gap}).

As usual, if one is interested in the vacuum state of the theory
(and the vacuum state preserves translational invariance), it is
possible to adopt, in place of the effective action
(\ref{gamma1pi}), the simpler effective potential $V(\varphi)$
defined by
\begin{equation}\label{veff}
    \Gamma[\varphi,G_{0}]\equiv -\int dx^{4} V(\varphi).
\end{equation}

 This CJT formalism which at first sight might appear more involved than the
standard formalism, with only the source $J$ coupled to the field,
turns out to be very useful in the so-called bubble resummation,
which takes into account all the vacuum to vacuum diagrams of the
type of Fig.\ref{fig:bd}. In the CJT formalism one obtains the
whole bubble resummation \cite{CJT} simply considering the
``eight''-diagram ($A$ in Fig.\ref{fig:vvdiagrams}) contribution
to $\Gamma^{2}(\varphi,G)$. We will use this approximation, also
known as Hartree-Fock approximation, in the following sections.

\section{CJT formalism in noncommutative field theories}

Having outlined the basic features of the CJT formalism we can
proceed to apply it to the noncommutative $\lambda\varphi^{4}$
scalar theory. We recall \cite{mrs}  that the action is given by
\begin{equation}
S=\int dx^{4}\left\{  \frac{1}{2}m^{2}\phi^{2}+\frac{1}{2}\partial_{\mu}%
\phi\partial^{\mu}\phi+\frac{\lambda}{4!}\phi\star\phi\star\phi\star
\phi\right\},  \label{clact}%
\end{equation}
where $\star$ represents the Moyal product:

\begin{equation}
\varphi_{1}(x)\star\varphi_{2}(x)=\varphi_{1}(x)e^{\frac{i}{2}\theta^{\mu\nu
}\overleftarrow{\partial}_{\mu}\overrightarrow{\partial}_{\nu}}\varphi_{2}(x),
\label{mp}%
\end{equation}
with $\theta^{\mu\nu}$ coordinate-independent antisymmetric
matrix.

In terms od the Moyal $\star$-product (\ref{mp}) the quadratic
part of the action remains unaffected by noncommutativity, while a
$\theta$-dependence arises for the interaction term. No specific
assumptions are made in the CJT formalism about the form of the
interaction therefore procedure outlined in Section-2 (and in
particular Eq.(\ref{gammae})) is still valid in this
noncommutative context.

It is easily seen that in the case under consideration:
\begin{align}
D^{-1}(x,y)  &  =\frac{\delta^{2}S}{\delta\varphi(x)\delta\varphi
(y)}=\label{invD}\\
&  =-\left[  \square+m^{2}\right]  _{x}\delta^{4}(x-y)-\frac{\lambda}%
{3!}\left\{
\delta^{4}(x-y)\star\varphi\star\varphi+\varphi\star\delta
^{4}(x-y)\star\varphi+\varphi\star\varphi\star\delta^{4}(x-y)\right\},
\nonumber\\
D_{0}^{-1}(x,y)  &  =D^{-1}|_{\lambda=0}=-\left[
\square+m^{2}\right]
_{x}\delta^{4}(x-y). \label{invD0}%
\end{align}

As expected $D_{0}^{-1}(x,y)$ is not modified by noncommutativity
since the integrals of terms quadratic in the fields are not
modified by the $\star$-product (\ref{mp}), while $D^{-1}(x,y)$
acquires the $\theta$-dependence. The $\star$-products which
appear in (\ref{invD}) can be easily calculated in momentum space.

Next let us consider the $S_{int}$ action. One can proceed in two
steps. The first step is the translation of the action
$S(\phi)\rightarrow S(\phi+\varphi)$. The second step is the one
of retaining from the shifted action only cubic, and higher, terms
in $\phi.$ So doing one obtains
\[
S_{int}(\phi;\varphi)=\frac{\lambda}{4!}\int
dx^{4}\phi\star\phi\star\phi \star\phi+\frac{\lambda}{6}\int
dx^{4}\phi\star\phi\star\phi\star\varphi,
\]
where the cyclicity of the $\star$-product (\ref{mp}) under
integration has been used.

To proceed further we now need to adopt an ansatz for the form of
$G(x,y).$ Since we intend to obtain the CJT effective potential we
must assume translational invariance, so that the more general
ansatz we can consider takes the form
\begin{equation}
G(x,y)=G(x-y)=\int \frac{d \alpha^{4}}{(2\pi)^{4}}\frac{\exp i\alpha(x-y)}{\alpha^{2}+M^{2}%
(\alpha)}\label{GA}
\end{equation}
where $M^{2}(\alpha)$ is to be determined. Once the ansatz
(\ref{GA}) has been done we can calculate the various terms in the
left-hand-side of (\ref{gammae}). The result is

\begin{align}
\Gamma(\varphi,G)  &  =\frac{1}{2}\int dx^{4}\left\{  (\partial_{\mu}%
\varphi)^{2}+m^{2}\varphi^{2}\right\}  +\frac{\lambda}{4!}\int dx^{4}%
\varphi\star\varphi\star\varphi\star\varphi+\label{GammaF}\\
&  +\frac{1}{2}\int dx^{4}\int \frac{dp^{4}}{(2\pi)^{4}}\ln\left\{  \frac{p^{2}+M^{2}(p)}%
{p^{2}+m^{2}}\right\}  +\nonumber\\
&  +\frac{1}{2}\int dx^{4}\int \frac{dk^{4}}{(2\pi)^{4}}\frac{m^{2}-M^{2}(k)+\dfrac{\lambda}%
{3!}\int \frac{dp_{1}^{4}dp_{2}^{4}}{(2\pi)^{8}} F(p_{1},p_{2},k)\exp ix(p_{1}+p_{2}%
)\widetilde{\varphi}(p_{1})\widetilde{\varphi}(p_{2})}{k^{2}+M^{2}%
(k)}+\nonumber\\
&  +\Gamma^{2}(\varphi,G)\nonumber
\end{align}
where $\widetilde{\varphi}(p)$ is the Fourier transform of
$\varphi(x)$. $F(p_{1},p_{2},k)$, using the notation $p_{i}\theta
p_{j} \equiv p_{i}^{\mu}\theta_{\mu \nu} p_{j}^{\nu}$, takes the
form
\begin{align}
 F(p_{1},p_{2},k)& =\exp -\frac{i}{2}\left(p_{1} \theta p_{2} +k
\theta p_{1} + k \theta p_{2} \right)+\nonumber\\
 & + \exp
-\frac{i}{2}\left(p_{1} \theta k +p_{1} \theta p_{2} + k \theta
p_{2} \right)+ \nonumber\\
 & +\exp -\frac{i}{2}\left(p_{1} \theta
p_{2} +p_{1} \theta k + p_{2} \theta k \right).
\end{align}
The first row in (\ref{GammaF}) is just $S(\varphi)$, the second
row is $Tr\ln(D_{0}^{-1}G)$, the third is $Tr\left\{
D^{-1}G-1\right\}$.

Now we must evaluate $\Gamma^{2}(\varphi,G)$ with vertices given
by $S_{int}(\phi;\varphi)$ and propagator given by $G(x,y).$ As a
first application of the formalism one can recover the usual
one-loop 1PI effective action \cite{mrs} simply setting
$\Gamma^{2}=0$. In this case the gap equation (\ref{gap2})
trivially reduces to
\begin{equation}
G^{-1}(x,y)=D^{-1}(x,y),
\end{equation}
and (\ref{gammae}) becomes
\[
\Gamma(\varphi)=S(\varphi)+\frac{1}{2}Tr\ln D^{-1}(\varphi)D_{0}.
\]
That is just the one-loop 1PI effective action.

While neglecting $\Gamma^{2}$ completely gives us back the
one-loop 1PI effective action, a more interesting result is
obtained by approximating $\Gamma^{2}(\varphi,G)$ including only
the ``eight'' diagram (A in Fig.\ref{fig:vvdiagrams}). In this
approximation one has that
\begin{equation}
\Gamma^{2}(\varphi,G)=\frac{1}{4!}\lambda\delta^{4}(0)\int \frac{d\alpha^{4}d\alpha^{\prime4}}{(2\pi)^{8}}%
\frac{1}{\alpha^{2}+M^{2}(\alpha)}\frac{1}{\alpha^{\prime
2}+M^{2}(\alpha^{\prime})}\left\{  1+2\cos^{2}(\frac{\alpha\theta
\alpha^{\prime}}{2})\right\}  . \label{G2}%
\end{equation}

We observe that differently from the commutative case, where the
momenta circulating in each of the two loops do not mix (so that
$\Gamma^{2}$ is evaluated as the products of two single-variable
integrals), in the commutative case this mixing occurs.

\section{The CJT effective potential in noncommutative field theories}

From the effective action (\ref{GammaF}) and (\ref{G2}), assuming
translational invariance of the candidate vacua $\varphi
(x)=\varphi$, one can extract the effective potential
(\ref{veff}):
\begin{align}
V(\varphi)  &
=\frac{1}{2}m^{2}\varphi^{2}+\frac{\lambda}{4!}\varphi
^{4}+\label{pp}\\
&  +\frac{1}{2}\int \frac{dp^{4}}{(2\pi)^{4}}    \ln\left\{  \frac{p^{2}+M^{2}(p)}{p^{2}+m^{2}%
}\right\}  +\nonumber\\
&  +\frac{1}{2}\int \frac{dk^{4}}{(2\pi)^{4}}\frac{m^{2}-M^{2}(k)+\dfrac{\lambda}{2}\varphi^{2}%
}{k^{2}+M^{2}(k)}+\nonumber\\
&  +\frac{1}{4!}\lambda\int
\frac{d\alpha^{4}d\alpha^{\prime4}}{(2\pi)^{8}} \frac{1}{\alpha
^{2}+M^{2}(\alpha)}\frac{1}{\alpha^{\prime2}+M^{2}(\alpha^{\prime})}\left\{
2+\cos(\alpha\theta\alpha^{\prime})\right\}  .\nonumber
\end{align}

The stationarity condition (\ref{gap}) in this case becomes
$\frac{\partial V(\varphi)}{\partial\varphi}=0$ which implies the
gap equation
\begin{equation}
M^{2}(\alpha)-m^{2}%
-\dfrac{\lambda}{2}\varphi^{2}-\frac{\lambda}{6}\int
\frac{db^{4}}{(2\pi)^{4}}  \frac{2+\cos\left(
b\theta\alpha\right)} {b^{2}+M^{2}(b)}=0. \label{gapquation}
\end{equation}

Substituting equation (\ref{gapquation}) in the expression of the
potential (\ref{pp}) we obtain
\begin{align}
V(\varphi,G)  &
=\frac{1}{2}m^{2}\varphi^{2}+\frac{\lambda}{4!}\varphi
^{4}+\frac{1}{2}\int \frac{dp^{4}}{(2\pi)^{4}}\ln\left\{  \frac{p^{2}+M^{2}(p)}{p^{2}+m^{2}%
}\right\}  +\label{potential}\\
&  -\frac{\lambda}{24}\int
\frac{dk^{4}}{(2\pi)^{4}}\frac{1}{k^{2}+M^{2}(k)}\int
\frac{db^{4}}{(2\pi)^{4}}\frac {1}{b^{2}+M^{2}(b)}\left\{
2+\cos\left(  b\theta k\right) \right\}. \nonumber
\end{align}

The term in the second row of the above expression is generated by
the bubble summation. The terms appearing in the first row are
already present at one-loop level but what is different here is
that they now must be evaluated for $M^{2}(p)$ solution of the gap
equation (\ref{gapquation}) which involves a non-trivial
$\theta$-dependence. Clearly our CJT resummed effective potential
has complicated $\theta$-dependence whereas the 1-loop effective
potential is completely $\theta$-independent. We notice that
$\theta$-dependence manifests in two different ways: the
tree-level and the one-loop like contributions depend on $\theta$
only through $M^{2}(p)$ whereas the term generated by the bubble
summation (the last one in (\ref{potential})) also exhibits an
explicit $\theta$-dependence.

Both (\ref{gapquation}) and (\ref{potential}) are ultraviolet
divergent and they both are considered to be regularized with a
cutoff $\Lambda$ on the loop-momenta. In the next section we will
deal with the problem of their renormalization but first we recall
the strategy of renormalization adopted in the commutative case
($\theta=0$).

\subsection{Renormalization of the effective potential for $\theta=0$}

In the commutative case (\ref{gapquation}) and (\ref{potential})
become respectively
\begin{equation}
M^{2}(\alpha)=m^{2}+\dfrac{\lambda}{2}\varphi^{2}+\frac{\lambda}{2}\int
\frac{db^{4}}{(2\pi)^{4}} \frac{1}{b^{2}+M^{2}(b)} \label{gcom}%
\end{equation}
and
\begin{equation}
V(\varphi,G)=\frac{1}{2}m^{2}\varphi^{2}+\frac{\lambda}{4!}\varphi^{4}%
+\frac{1}{2}\int \frac{dp^{4}}{(2\pi)^{4}}\ln\left\{
\frac{p^{2}+M^{2}(p)}{p^{2}+m^{2}}\right\} -\frac{\lambda}{8}\int
\frac{dk^{4}}{(2\pi)^{4}}\frac{1}{k^{2}+M^{2}(k)}\int
\frac{db^{4}}{(2\pi)^{4}}\frac{1}
{b^{2}+M^{2}(b)}. \label{vcom}%
\end{equation}

We immediately see, from Eq.(\ref{gcom}), that in this case
$M^{2}(\alpha)$ does not depend on the loop-momentum $\alpha$ so
that one can set $M^{2}(\alpha)\equiv M^{2}$ both in (\ref{gcom})
and in (\ref{vcom}).

Renormalization of the effective potential for $\theta=0$ is a
well-known result (see e.g. \cite{gacpi}). We recall the procedure
that one can use to renormalize (\ref{gcom}) and (\ref{vcom})
since we will use it widely in the rest of the paper. In order to
renormalize this gap equation (\ref{gcom}) it is convenient to
introduce $I_{1},I_{2},G_{R}$ through the formula
\begin{equation}
    \int \frac{db^{4}}{(2\pi)^{4}}\frac{1}{b^{2}+M^{2}}=I_{1}-I_{2}M^{2}+G_{R}(M),
\end{equation}
where
\begin{align}
 I_{1}& =\int \frac{db^{4}}{(2\pi)^{4}}\frac {1}{b^{2}}, \\
I_{2}& =\int \frac{db^{4}}{(2\pi)^{4}}\frac{1}{b^{4}},
\end{align}
and $G_{R}(M)$ is the finite part of $G(M)$. The gap equation can
then be rewritten in the form
\begin{equation}
M^{2}\left(\frac{1}{\lambda} + \frac{1}{2} I_{2}
\right)=\frac{m^{2}}{\lambda} +
\frac{1}{2}I_{1}+\frac{1}{2}\varphi^{2}+\frac{1}{3} G_{R}(M).
\end{equation}

Introducing the renormalized parameters
\begin{align}
\dfrac{1}{\lambda_{R}} &  =\lim_{\Lambda\rightarrow\infty}\left[
\frac
{1}{\lambda}+\frac{1}{2}I_{1}\right]  ,\label{lrc}\\
\dfrac{m_{R}^{2}}{\lambda_{R}} &
=\lim_{\Lambda\rightarrow\infty}\left[
\frac{m^{2}}{\lambda}+\frac{1}{2}I_{2}\right]  ,\label{mrc}%
\end{align}
one gets the renormalized gap equation
\[
M^{2}=m_{R}^{2}+\dfrac{\lambda_{R}}{2}\varphi^{2}+\frac{\lambda_{R}}{3}%
G_{R}(M).
\]
Next one verifies that the conditions (\ref{lrc}) and (\ref{mrc})
also renormalize the effective potential. For this objective one
observes that using (\ref{gcom}) one can write (\ref{vcom}) as the
sum of three contributions
\begin{align}
V^{0} &  =\frac{1}{2}m^{2}\varphi^{2}+\frac{\lambda}{4!}\varphi^{4}%
,\label{vc1}\\
V^{I} &  =\frac{1}{2}\int \frac{dk^{4}}{(2\pi)^{4}}\ln(k^{2})+\frac{1}{2}I_{1}M^{2}-\frac{1}%
{4}I_{2}M^{4}+T(M)\label{vc2}\\
V^{II} &  =-\frac{1}{2}GM^{2}+\frac{1}{2\lambda}M^{4}-\frac{1}{2\lambda}%
m^{4}-\frac{1}{2}m^{2}\varphi^{2}-\frac{\lambda}{8}\varphi^{4},\label{vc3}%
\end{align}
that gives
\begin{equation}
V=V^{II}+V^{I}+V^{0}=-\frac{\lambda}{12}\varphi^{4}+\frac{1}{2}\frac{M^{4}%
}{\lambda_{R}}-\frac{1}{2}M^{2}G_{R}(M)+T(M),\label{vcr}%
\end{equation}
where $T(M)$ is a finite contribution to $V^{I}$ (which of course
is not significant for renormalizability) and the
field-independent terms have been discarded.

We observe that the CJT effective potential (\ref{vcr}) in terms
of the renormalized parameters, defined by (\ref{lrc}) and
(\ref{mrc}), is finite. One could be concerned with the
unrenormalized contribution $-\frac{\lambda}{12}\varphi^{4}$ but
since from Eq.(\ref{lrc}) it follows that
$\lambda(\Lambda)\simeq(\lambda_{R}-\frac{1}{2}I_{2})^{-1}$ the
term $-\frac{\lambda}{12}\varphi^{4}$ vanishes upon removal of the
cutoff.\footnote{Actually as emphasized in the literature (see
e.g.\cite{gacmf}), for $\Lambda \rightarrow \infty$ with
fixed-positive nonzero $\lambda_{R}$ one finds from Eq.(\ref{lrc})
that $\lambda$ is vanishingly small but negative ($\lambda
\rightarrow 0^{-}$). This non-positive value of the bare coupling
can be concerning \cite{gacmf} for the stability of the theory.
However it simply reflects the well known triviality of $\lambda
\varphi^{4}$ theory (see e.g.\cite{tst}) and is usually handled by
considering a large but finite cutoff.}

In preparation for a key observation made in the following section
let us also emphasize the exact cancellation of the potentially
dangerous divergent terms of type $m^{2}\varphi^{2}$ which appear
with opposite signs in the tree level contribution $V^{0}$ and in
the loop correction $V^{II}$.

\subsection{Renormalization of the effective potential for $\theta\neq0$}

 Now we want to address the problem of evaluating the effective potential for nonvanishing
 values of the noncommutativity parameter $\theta$. In the analysis
of the gap equation (\ref{gapquation}) one finds immediately an
important complication due to nonvanishing $\theta$. In fact a
constant $M^{2}$ clearly cannot be a solution of
(\ref{gapquation}) and the gap equation must then be solved for a
momentum-dependent function $M^{2}(p)$.

We start by defining $M^{2}%
(\alpha)=M^{2}+\Pi(\alpha)$ so that the gap equation
(\ref{gapquation}) can be rewritten as
\[
\Pi(\alpha)=-M^{2}+m^{2}+\dfrac{\lambda}{2}\varphi^{2}+\frac{\lambda}{3}\int
\frac{db^{4}}{(2\pi)^{4}}\frac{1}{b^{2}+M^{2}+\Pi(b)}+\frac{\lambda}{6}\int
\frac{db^{4}}{(2\pi)^{4}}\frac {\cos\left(  b\theta\alpha\right)
}{b^{2}+M^{2}+\Pi(b)}.
\]

We use the freedom in the choice of the ``constant'' part of
$M^{2}(\alpha)$ to make it satisfy

\begin{equation}
M^{2}=m^{2}+\dfrac{\lambda}{2}\varphi^{2}+\frac{\lambda}{3}\int
\frac{db^{4}}{(2\pi)^{4}}
\frac{1}{b^{2}+M^{2}+\Pi(b)},\label{pieo}%
\end{equation}
and

\begin{equation}
\Pi(\alpha)=\frac{\lambda}{6}\int
\frac{db^{4}}{(2\pi)^{4}}\frac{\cos\left(  b\theta \alpha\right)
}{b^{2}+M^{2}+\Pi(b)}. \label{pie}
\end{equation}

We notice that the integral in the right-hand-side of
Eq.(\ref{pie}) is finite both in the ultraviolet sector and in the
infrared sector so that Eq.(\ref{pie}) does not need to be
renormalized. We also observe that $\Pi(\alpha)\rightarrow\infty$
for $\alpha \rightarrow0$ and for $\theta\rightarrow0$, and that
$\Pi(\alpha)\rightarrow0$ for $\alpha\rightarrow\infty$ and for
$\theta\rightarrow\infty$.

Equation (\ref{pieo}), on the contrary, is not UV-finite. It can
be renormalized by a procedure similar to the one we have
discussed previously. Introducing the renormalized parameters in
the form
\begin{align}
\dfrac{1}{\lambda_{R}}  & =\lim_{\Lambda\rightarrow\infty}
 \left[\frac{1}{\lambda
}+\frac{1}{3}I_{2}\right],\label{rc}\\
\dfrac{m_{R}^{2}}{\lambda_{R}}  &
=\lim_{\Lambda\rightarrow\infty}\left[\frac
{m^{2}}{\lambda}+\frac{1}{3}I_{1}\right], \label{rm}%
\end{align}
we get the renormalized gap equation
\[
M^{2}=m_{R}^{2}+\dfrac{\lambda_{R}}{2}\varphi^{2}+\frac{\lambda_{R}}{3}
F_{R}(M),
\]
where $F_{R}(M)$ is the finite part of the divergent integral in
(\ref{pieo}):
\begin{equation}
    F_{R}(M)=\int
\frac{db^{4}}{(2\pi)^{4}} \frac{1}{b^{2}+M^{2}+\Pi(b)}-\left[\int
\frac{db^{4}}{(2\pi)^{4}} \frac{1}{b^{2}}-M^{2}\int
\frac{db^{4}}{(2\pi)^{4}} \frac{1}{b^{4}}\right]
\end{equation}

We come now to the important issue of the renormalization of the
effective potential. We have seen that the way in which the gap
equation renormalizes fixes uniquely the renormalization of the
bare mass and the renormalization of the coupling. We must check
if the same renormalization conditions provide us a finite
effective potential. We can use (\ref{pieo}) and (\ref{pie}) in
the expression (\ref{potential}) to obtain the effective potential
as the sum of the three following terms:
\begin{align}
V^{0} &  =\frac{1}{2}m^{2}\varphi^{2}+\frac{\lambda}{4!}\varphi^{4}%
,\label{v1nc}\\
V^{I} &  =\frac{1}{2}\int \frac{dk^{4}}{(2\pi)^{4}}\ln(k^{2})+\frac{1}{2}I_{1}M^{2}-\frac{1}%
{4}I_{2}M^{4}+T(M,\Pi),\label{v2nc}\\
V^{II} &  =M^{4}\left\{\frac{1}{2}I_{2}+\dfrac{3}{4}\dfrac{1}{\lambda}\right\}+
\frac{M^{2}}{2}\int\frac{dk^{4}}{(2\pi)^{4}}\frac{\Pi(k)[\Pi(k)+2k^{2}]}{k^{4}\left[k^{2}+\Pi(k)\right]^{2}}   \nonumber\\
&  +M^{2}\left\{-\dfrac{1}{2}I_{1}  -\dfrac{1}{2}\int
\frac{dk^{4}}{(2\pi)^{4}}R(k)+\dfrac{1}{4}\int \frac{dk^{4}}{(2\pi)^{4}}\frac{\Pi(k)}{ k^{2}%
+\Pi(k)} \left( \frac{1}{k^{2}+\Pi(k)}+\frac{2}{k^{2}}  \right)       \right\}  +\nonumber\\
&  -\dfrac{1}{4}\int
\frac{dk^{4}}{(2\pi)^{4}}\frac{\Pi(k)}{k^{2}+\Pi(k)}-\dfrac{1}{4}\int
\frac{dk^{4}}{(2\pi)^{4}}\Pi(k)R(k)-\dfrac{3m^{4}}{4\lambda}-\dfrac{3}{4}m^{2}\varphi^{2}%
-\dfrac{3}{16}\lambda\varphi^{4},\label{v3nc}%
\end{align}
where we  have defined
\[
T(M,\Pi)=\dfrac{1}{2}\int
\frac{dk^{4}}{(2\pi)^{4}}\ln[k^{2}+M^{2}+\Pi(k)]-\left\{
\dfrac{1}{2}\int
\frac{dk^{4}}{(2\pi)^{4}}\ln[k^{2}]+\frac{M^{2}}{2}\int
\frac{dk^{4}}{(2\pi)^{4}}\frac{1}{k^{2}}-\frac{M^{4}}{4}\int
\frac{dk^{4}}{(2\pi)^{4}}\frac{1}{k^{4} }\right\},
\]
and
\begin{equation}
R(k)=\tfrac{1}{k^{2}+M^{2}+\Pi(k)}-\left\{  \tfrac{1}{k^{2}+\Pi(k)}%
-\tfrac{M^{2}}{\left[  k^{2}+\Pi(k)\right]  ^{2}}\right\}.
\end{equation}

We observe that in our CJT effective potential
($V=V^{0}+V^{I}+V^{II}$), thanks to the fact that $\Pi(k)$
vanishes exponentially in the limit $k\rightarrow\infty$, all the
field-dependent terms, with the exception of $m^{2}\varphi^{2}$
and $\lambda\varphi^{4}$ terms, are cutoff independent. The term
in $\lambda\varphi^{4}$ is not divergent, it is similar to the
$\lambda\varphi^{4}$-term present in the commutative case and can
be handled in the same way. The term in $m^{2}\varphi^{2}$, on the
contrary, is genuinely ultraviolet divergent. This is due to the
fact that the corresponding contributions from $V^{0}$ and
$V^{II}$ do not cancel each other (unlike the commutative case).
This clearly originates from the well-known fact that nonvanishing
$\theta$, by regularizing certain otherwise divergent diagrams,
effectively changes the numerical coefficients in front of some
divergent terms.

The presence of this $\varphi$-dependent contribution that
diverges upon cutoff removal leads to conjecture that it is
impossible for this theory to be in an ordered translationally
invariant phase\footnote{The fact that for $\varphi=0$ this
alarming divergence disappears might mean that instead a
translationally invariant disordered phase is possible.}.

\subsection{Renormalization of the effective potential in the planar limit ($\theta \rightarrow \infty$)}

 It is also interesting to consider the limit of strong
noncommutativity ($\theta\rightarrow\infty$). In this limit the
strong oscillations in the phases, which are present in the
integrands of (\ref{gapquation}) and (\ref{potential}), induce the
vanishing of the corresponding integrals. In this limit the
nonplanar contribution of the eight diagram becomes negligible and
only the planar contribution survives. In this strong
commutativity limit the effective potential is given by the sum of
the terms
\begin{align}
V^{0} &  =\frac{1}{2}m^{2}\varphi^{2}+\frac{\lambda}{4!}\varphi^{4},%
\label{v1snc}\\
V^{I} &  =\frac{1}{2}\int \frac{dk^{4}}{(2\pi)^{4}}\ln(k^{2})+\frac{1}{2}I_{1}M^{2}-\frac{1}%
{4}I_{2}M^{4}+T(M,0),\label{v2snc}\\
V^{II} &
=-\frac{3}{4\lambda}m^{4}-\dfrac{3\lambda}{16}\varphi^{4}+\frac
{3}{4\lambda}M^{4}-\dfrac{1}{2}M^{2}G(M)-\frac{3}{4}m^{2}\varphi^{2},%
\label{v3snc}%
\end{align}
so that one finds
\begin{equation}\
    V=\frac{3}{4}\frac{M^{4}}{\lambda_{R}}-\frac{7}{48}\lambda
    \varphi^{4}-\frac{1}{4}m^{2}\varphi^{2}-\dfrac{1}{2}M^{2}G_{R}(M)+T(M,0).
\end{equation}

This expressions for the effective potential is formally similar
to the ones of the commutative case, (\ref{vcr}). The crucial
differences consist in the factors in front to the eight-diagram
terms. It particular in this limit, as in the commutative case,
the dressed mass $M^{2}(k)=M^{2}+\Pi(k)$ becomes momentum
independent since $\Pi(k)$ vanishes exponentially.

Again the cancellation between the $m^{2}\varphi^{2}$ terms in
(\ref{v1snc}) and (\ref{v3snc}) doesn't occur and the resulting
potential doesn't renormalize.

\section{Conclusions}

\qquad Whereas in commutative spacetime the CJT effective
potential can be renormalized and gives a satisfactory description
of the vacua of a given field theory, in our
canonical-noncommutativity analysis the CJT effective potential
(in the bubble-resummation approximation) was found not to be
renormalizable. From a conservative standpoint we should then
assume that in this type of theories the CJT effective potential
cannot provide reliable nonperturbative insight on the phase
structure. This negative conclusion is also supported by the
realization that canonical noncommutativity affects strongly the
structure of the UV divergences of a field theory, and this might
be particularly significant for those techniques that effectively
rely on resummations of contributions from all orders in the
coupling constant. When we establish that a field theory is
renormalizable, we actually verify that it is ``perturbatively
renormalizable'': the divergences at any given order in
coupling-expansion perturbation theory can be reabsorbed in
redefinitions of the parameters of the Lagrangian density. The
fact that the CJT technique gives rise to a renormalizable
effective potential in the commutative-spacetime case is highly
nontrivial, since we are not consistently summing all
contributions up to a given order in the coupling constant (a
calculation which would be ``protected'' by perturbative
renormalizability), we are instead selectively summing a certain
subset of the contributions at all orders in the coupling
constant. It is therefore plausible that the fact that our CJT
effective potential cannot be renormalized is simply a sign of an
inadequacy of this technique to the canonical-noncommutativity
context.

On the other hand it appears reasonable to explore an alternative,
more optimistic, perspective, which is based on the observation
that the only contribution to the CJT potential that ends up not
being expressed in terms of renormalized quantities is the
$\frac{1}{4}m^{2}\varphi^{2}$ contribution. This term however will
vanish in a disordered phase $\varphi=0$. In a certain sense we
have a renormalizable effective potential in the disordered phase,
and our results of nonrenormalizability in the
translationally-invariant ordered phase $\varphi=C$ could be
interpreted as a manifestation of the fact that this phase is not
admissable for these theories in canonical noncommutative
spacetime. This hypothesis finds some support in the arguments
presented in Ref. \cite{guso}, which also concluded that the only
admissible phases for these theories are the disordered phase and
a
(non-translationally-invariant) stripe phase with $\tilde{\varphi}%
(p)=C\delta(p-p_{c})$ (where $\tilde{\varphi}(p)$\ is the Fourier
transform of $\varphi(x)$) and $p_{c}$ is a characteristic
momentum scale of the stripe phase). This argument of
inadmissability of the translationally-invariant ordered phase
might be related with the delicate IR structure of these theories:
$\varphi(x)=C$ means $\tilde{\varphi}(p)=C\delta(p)$, so the
concept of a translationally-invariant ordered phase is closely
connected with the zero-momentum structure of the theory of
interest.

The fact that various lines of analysis appear to suggest that a
non-translationally-invariant ordered phase should naturally be
considered in these theories is rather intriguing. In fact
canonical noncommutativity is naturally in conflict with
invariance under space rotations and/or boosts, but, when analyzed
as a pure algebra of coordinates, appears to be perfectly
compatible with classical translational invariance. If a non
translational invariant ordered phase emerges from field theories
in canonical noncommutative spacetime it would mean that these
theories might favour spontaneous breaking of their only residual
spacetime symmetry\footnote{It might be interesting to investigate
the possibility of analogous spontaneous symmetry-breaking
mechanism in noncommutative geometries such as the ones considered
in \cite{MR,LRZ,gacma} which are instead naturally invariant under
space rotations and (deformed \cite{dsr1,BAK}) boosts.} (the one
under spacetime translations).

To explore these issues it would be necessary to consider the CJT
effective action, which explores the more general class of
candidate vacua $\varphi(x)$, rather than stopping, as we did
here, at the level of the CJT effective potential (which assumes
from the beginning a translationally-invariant vacuum). With the
CJT effective action one could investigate the renormalizability
of the stripe phase (which is not translationally invariant, and
therefore cannot be studied with the effective potential).
Unfortunately even in commutative-spacetime theories the
evaluation of the CJT effective action turns out to be very
complex (basically intractable analytically and a troublesome
calculation even numerically). It is likely that in the
canonical-noncommutativity context the evaluation of the CJT
effective action may prove even more troublesome, but from the
indications that emerged from our analysis of the CJT effective
potential it appears that such an analysis is well motivated, as
it could provide insight for the understanding of some key
physical predictions of these theories.

\section*{Note added}
These results were first presented in Chapter 5, of my Ph.D.
thesis \cite{PT}. As I was rewriting the discussion in a format
more suitable for the article-style presentation here provided the
study in Ref.\cite{CZ} was announced. The results reported in
Ref.\cite{CZ} are consistent (and partially overlap) with the ones
presented here and in Ref.\cite{PT}.

\section*{Acknowledgements}
I thank Giovanni Amelino-Camelia for the constant support during
the whole period of preparation of this paper, for the
enlightening discussions, for bringing to my attention
Ref.\cite{CZ} and for the careful reading of the manuscript. I
also thank Luisa Doplicher for valuable suggestions and
stimulating comments.

\end{document}